\documentclass{ws-procs975x65}

\begin{document}

\title{The Physics of the Far Future}

\author{Ruxandra Bondarescu $^*$} 
\address{Institute for Theoretical Physics, Winterthurerstrasse 190, 
CH-8057, Zurich, Switzerland \\
E-mail: ruxandra@physik.uzh.ch}
\author{Andrew P. Lundgren}
\address{Albert Einstein Institut, Callinstr. 38,  
30167 Hannover, Germany\\
E-mail: andrew.lundgren@LIGO.org}
\author{Mihai Bondarescu}
\address{108 Lewis Hall, University of Mississippi, Oxford, MS 38677, USA}
\address{Facultatea de Fizica, Universitatea de Vest, 
Blvd.~V.~Parvan 4, Timisoara 300223, Romania \\
E-mail: mihai7@gmail.com}
\begin{abstract}
We observe the past and present of the universe, but can we predict the far future? Observations suggest that in thousands of billions of years from now most matter and radiation will be absorbed by the cosmological horizon. As it absorbs the contents of the universe, the cosmological horizon is pushed further and further away.  In time, the universe asymptotes towards an equilibrium state of the gravitational field. Flat Minkowski space is the limit of this process. It is indistinguishable from a space with an extremely small cosmological constant ($\Lambda \to 0$) and thus has divergent entropy.
\end{abstract}

\section{Introduction}
If we ignored gravitational entropy, we would think that the early universe had high entropy 
because it was uniform and hot \cite{ref1}, and that the cold and clumpy universe of today has low
entropy. The evolution of the universe would then violate the second law of thermodynamics, which
states that entropy never decreases. However, gravitational entropy cannot be ignored because it increases with clumpiness and ends up dominating the other forms of entropy. 

Gravitational entropy is usually identified with a quarter the area of a horizon, which suggests that the entropy arises in some way from the information that the horizon hides from us. In the case of the cosmological horizon, the expansion of the universe carries the distant object beyond contact with observers from within the horizon.  In the black hole case, the composition of an object is forgotten once it falls into the hole, and only a few macroscopic variables - mass, spin, and charge - remain. It would  then be easy conclude that gravitational entropy exists only in the presence of the horizon and is proportional to the information hidden by the horizon.  However, there are many counter-examples that suggest that gravitational entropy is non-zero even when a horizon is absent.  The gravitational-wave equivalent of the cosmic microwave background should have a temperature of about a Kelvin, and should have an entropy roughly in line with that of the CMB. Does a collapsing star, which is  $\epsilon$ away from forming a black hole and does not have a horizon yet, have gravitational entropy?  

Black holes are the most lasting objects in the universe. They can be fully described by three observables - mass, spin and charge and thus have a huge number of indistinguishable configurations.  In general, entropy is proportional to the number of indistinguishable configurations in a system. As a gas cloud collapses to form a star, and eventually forms a black hole, entropy increases.  It is not surprising then that the three largest repositories of entropy in the universe are gravitational: the cosmological horizon, the supermassive black holes and the cosmic microwave background. 

\section{The ``Closer" Far Future}
The local supercluster is gravitationally bound and is expected to collapse to a supermassive black hole \cite{ref8} with a Schwarzchild radius of $300$ light years and a mass $M \sim 10^{15} M_\odot$ \cite{ref16}. The Hawking temperature of this black hole would be $T_{\rm BH} \approx 1/(8 \pi M) \sim  10^{-23}$ K \cite{ref17, ref10}. In $10^{12}$ years when the CMB cools below $10^{-23}$ K, this black hole will start to evaporate. Heat will flow from the hotter black hole to the colder cosmological horizon producing much more entropy that was lost by the black hole. As the black hole evaporates, the total entropy increases by a factor proportional to $M \alpha$, which is much larger than the entropy of the black hole itself:
\begin{equation}
S_{\rm tot} \approx S_C = \pi (\alpha^2 - 2 \alpha M  + ....),
\end{equation}
where $\alpha = \sqrt{3/\Lambda}$ \cite{refour,ref21}. 

In due time, nothing will be left other than slowly evaporating supermassive black holes making black hole thermodynamics the physics of the far future! The evaporation process satisfies the second law of thermodynamics since the heat flows from hot (black hole) to cold (cosmological horizon) generating entropy.

\section{Predictions for the End State of the Universe} 
The emptying of the universe is its approach to equilibrium \cite{ref26}. The state of maximum entropy of a gas in a box is its equilibrium state. It is also the simplest state the gas can be in, i.e., uniformly distributed and featureless. In the case of the universe, the simplest, most uniform state is flat space. Empty deSitter is maximally symmetric.  It posses a full set of ten symmetries expressed by Killing vectors (one time and three space translations, three rotations, and three boosts). However, it has a preferred scale given by the cosmological constant $\Lambda$ \cite{ref27,ref28}, which suppresses very long wavelength modes of the Weyl tensor also known as gravitational waves. Flat space Minknowski space therefore has even higher entropy than deSitter because it has the same symmetries,  but no suppression factor. 

Flat Minkowski can be thought of as deSitter with an extremely small cosmological constant $\Lambda \to 0$. In this limit, the entropy is divergent \cite{ref20}, which introduces an unregulated infinity in the calculation when we use flat space as a background in black hole thermodynamics. For this reason, we propose that deSitter space-time is a better background than Minkowski. 

\section{Discussion and Conclusions}
 Gravitational entropy has to taken into account to explain the evolution of the universe. We do not know how to compute gravitational entropy in general situations when no horizon is present. A first attempt to compute gravitational entropy inside a star from Brown-York \cite{BrownYork} quasi-local quantities is performed by Lundgren {\it et al.}.\ \cite{ustwo}. 
 
Flat space has higher entropy than deSitter space. The second law of thermodynamics therefore favors the decay of deSitter to flat space, which may happen when quantum effects are included \cite{refChair}. Alternatively, if the cosmological constant is mimicked by a scalar field, our universe could cascade through a series of configurations with increasing entropy and decreasing apparent dark energy density, and eventually reach flat space.

\bodymatter

\end{document}